\documentclass[12pt]{article}
\usepackage{graphicx}
\usepackage{amsmath}
\usepackage{amssymb}
\usepackage{caption2}
\begin{document}
\bibliographystyle{prsty}
\begin{center}
{\large {\bf \sc{ Analysis the $f_0(980)$ and $a_0(980)$ mesons
as four-quark states with  the QCD sum rules }}} \\[2mm]
Zhi-Gang Wang$^{1}$ \footnote{Corresponding author;
E-mail,wangzgyiti@yahoo.com.cn. } and
Wei-Min Yang$^{2}$   \\
$^{1}$ Department of Physics, North China Electric Power University, Baoding 071003, P. R. China \\
$^{2}$ Department of Modern Physics, University of Science and Technology of China, Hefei 230026, P. R. China \\
\end{center}

\begin{abstract}
In this article, we take the point of view  that the scalar mesons
$ f_0(980)$ and $a_0(980)$ are diquark-antidiquark states
$(qq)_{\bar{3}}(\bar{q}\bar{q})_3$, and devote to determine their
masses in the framework of the QCD sum rules approach with the
interpolating currents constructed from scalar-scalar type and
pseudoscalar-pseudoscalar type diquark pairs respectively. The
numerical results indicate that the scalar mesons $ f_0(980)$ and
$a_0(980)$ may have two possible diquark-antidiquark
substructures.
\end{abstract}


\section{Introduction}
The light flavor  scalar mesons present a remarkable exception for
the constituent  quark model and the structures of those mesons
have not been unambiguously determined yet \cite{Godfray}.
Experimentally, the strong overlaps with each other
 and the broad widths ( for the $f_0(980)$, $a_0(980)$ et al, the widths are comparatively narrow)
 make their spectra cannot be approximated by the
  Breit-Wigner  formula.  The numerous
candidates  with the same quantum numbers $J^{PC}=0^{++}$ below $2
GeV$ can not be accommodated in one $q\bar{q}$ nonet,  some are
supposed to be glueballs, molecules and multiquark states. The
more elusive things are the constituent structures of the mesons
$f_0(980)$ and $a_0(980)$  with almost the degenerate masses.  In
the naive  quark model,
$a_0=(u\overline{u}-d\overline{d})/\sqrt{2}$ and
$f_0=s\overline{s}$; while in the framework of  four-quark
  models,  the  mesons $f_0(980)$ and $a_0(980)$
could either
 be compact  objects i.e. nucleon-like bound states of
quarks  with symbolic quark structures  $f_0={s{\overline s}({ u
{\overline u}+d {\overline d})/ \sqrt{2}}}$ and $a_0=s {\overline
s}( u {\overline u}-d {\overline d}) / \sqrt{2}$
\cite{JaffeAchasov4q}, or spatially  extended objects i.e.
deuteron-like bound states of hadrons,  the $f_0(980)$ and
$a_0(980)$ mesons are usually taken as $K {\overline K}$ molecules
\cite{IsgurKK}. The hadronic dressing mechanism takes the point of
view that the mesons  $f_0(980)$ and $a_0(980)$ have small
$q\bar{q}$ cores of typical $q\bar{q}$ meson size, the strong
couplings to the hadronic channels  enrich the pure $q\bar{q}$
states with other components and spend part (or most part) of
their lifetime as virtual $ K \bar{K} $ states \cite{HDress}. In
the hybrid model, those mesons are four-quark states
$(qq)_{\bar{3}}(\bar{q}\bar{q})_3$  in S-wave near the center,
with some constituent  $q \bar{ q}$ in P-wave, but further out
they rearrange into  $(q \bar{ q})_1(q \bar{ q})_1$ states and
finally as meson-meson states \cite{Close2002}.
  All those interpretations have both outstanding
advantages and obvious shortcomings in one or other ways.

There maybe exist two scalar nonets   below $1.7 GeV$. The
attractive interactions of one gluon exchange  favor the formation
of diquarks in  color antitriplet $\overline{3}_{ c}$, flavor
antitriplet $\overline{3}_{ f}$ and spin singlet  $ 1_{s} $.  The
strong attractions between the states $(qq)_{\overline{3}}$ and
$(\bar{q}\bar{q})_{3}$  in S-wave may result in a nonet manifested
below $1GeV$ while the conventional $^3P_0$ $\bar{q}q$ nonet would
have masses about $1.2-1.6 GeV$. Taking  the diquarks and
antidiquarks as the basic constituents, keeping  the effects of
the $s$ quark mass at the first order,  the two isoscalars $\bar
u\bar d u d$ and $\bar s s\frac{\bar u u+\bar d d}{\sqrt{2}}$ mix
ideally, the $\bar s s\frac{\bar u u+\bar d d}{\sqrt{2}}$
 degenerate with the isovectors  $\bar s s\bar d u$,
$\bar s s\frac{\bar u u-\bar d d}{\sqrt{2}}$ and $\bar s s\bar u
d$ naturally. Comparing with  the traditional ${\bar q} q $ nonet
mesons, the mass spectrum is inverted.  The lightest state is the
non-strange isosinglet ($\bar u\bar d ud$), the heaviest are the
degenerate isosinglet and isovectors with hidden $\bar s s$ pairs
while the four strange states lie in between
\cite{Close2002,ReviewScalar}.

In this article, we take the point of view that the well confirmed
$f_0(980)$ and $a_0(980)$ mesons are  four-quark states
$(qq)_{\bar{3}}(\bar{q}\bar{q})_3$ in the ideal mixing limit, and
devote to determine the values of their masses $m_{f_0}$ and
$m_{a_{0}}$ in the framework of  the  QCD sum rules approach
\cite{Shifman79,Maiani04,Nielsen04}. Detailed studies about the
other  scalar four-quark states ( the $\kappa(800)$s are not
confirmed yet) and the mixing between the two isoscalars (the
$f_0(980)$ meson and the broad $f_0(600)$ meson) will be our next
work.

The article is arranged as follows: in section II, we obtain the
QCD sum rules for the masses of the mesons $f_0(980)$ and $a_0(980)$
; in section III, numerical results; section IV is reserved
for conclusion.

\section{ Masses of the $f_0(980)$ and
$a_0(980)$ mesons  with the  QCD Sum Rules}

In the four-quark models, the structures of the scalar mesons
$f_0(980)$ and $a_0(980)$ in the ideal mixing limit can be
symbolically taken as \cite{JaffeAchasov4q,Close2002,ReviewScalar}
\begin{eqnarray}
f_0(980)=\frac{us\bar{u}\bar{s}+ ds\bar{d}\bar{s}}{\sqrt{2}},
\,\,\, a_0(980)=\frac{us\bar{u}\bar{s}-
ds\bar{d}\bar{s}}{\sqrt{2}}.
\end{eqnarray}
The four-quark configurations of the  $J^{PC}=0^{++}$ mesons can
give a lot of satisfactory  descriptions of the hadron phenomenon,
for example, the mass degeneracy of the $f_0(980)$ and $a_0(980)$
mesons, the mass hierarchy pattern of the scalar nonet, the large
radiative widths of the $f_0(980)$ and $a_0(980)$ mesons, the
$D_s^+(c\bar{s})$ to $\pi^+\pi^+\pi^-$ decay.

In the following, we write down the interpolating currents  for
the scalar mesons $f_0(980)$ and $a_0(980)$ based on the
four-quark model \cite{Maiani04,Nielsen04},
\begin{eqnarray}
J^A_{f_0}&=&\frac{\epsilon_{abc}\epsilon_{ade}}{\sqrt{2}}\left[(u_b^TC\gamma_5s_c)
(\bar{u}_d\gamma_5C\bar{s}_e^T)+(d_b^TC\gamma_5s_c)
(\bar{d}_d\gamma_5C\bar{s}_e^T)\right], \\
J^B_{f_0}&=&\frac{\epsilon_{abc}\epsilon_{ade}}{\sqrt{2}}\left[(u_b^TCs_c)
(\bar{u}_dC\bar{s}_e^T)+(d_b^TC s_c) (\bar{d}_d
C\bar{s}_e^T)\right], \\
J^A_{a_0}&=&\frac{\epsilon_{abc}\epsilon_{ade}}{\sqrt{2}}\left[(u_b^TC\gamma_5s_c)
(\bar{u}_d\gamma_5C\bar{s}_e^T)-(d_b^TC\gamma_5s_c)
(\bar{d}_d\gamma_5C\bar{s}_e^T)\right], \\
J^B_{a_0}&=&\frac{\epsilon_{abc}\epsilon_{ade}}{\sqrt{2}}\left[(u_b^TCs_c)
(\bar{u}_dC\bar{s}_e^T)-(d_b^TCs_c)
(\bar{d}_dC\bar{s}_e^T)\right],
\end{eqnarray}
where $a,~b,~c,~...$ are color indices and $C$ is the charge
conjugation matrix.  The constituents $S^a(x) = \epsilon^{abc}
u_b^T(x)C\gamma_5 s_c(x) $ and $P^a(x) = \epsilon^{abc}
u_b^T(x)Cs_c(x)  $ represent the scalar $J^P=0^+$ and the
pseudoscalar $J^P=0^-$ $us $ diquarks respectively. They both
belong to the antitriplet $\bar{3}$ representation of the color
$SU(3)$ group and can cluster  together to form    $S^a-\bar{S}^a$
type and $P^a-\bar{P}^a$ type diquarks pairs to give the correct
spin and parity for the scalar mesons $J^P=0^+$ . The scalar
diquarks correspond to the $^1S_0$ states of $us$ and $ds$ diquark
systems. The one gluon exchange force and the instanton induced
force can lead to significant attractions between the quarks in
the $0^+$ channels \cite{GluonInstanton}. The pseudoscalar
diquarks do not have nonrelativistic limit,  can be  taken as  the
$^3P_0$ states.

  The calculation of the  $a_0(980)$ meson as a four-quark
state in the QCD sum rules approach  was done  originally for the
 decay constant and the hadronic coupling constants
with the interpolating currents $J_{a_0}^1$ and $J_{a_0}^2$
\cite{Latorre85,Narison86},
\begin{eqnarray}
J_{f_0(a_0)}^1&=&\Sigma_{\Gamma=1,\pm \gamma_5} \bar{s}\Gamma
s\frac{\bar{u}\Gamma u\pm \bar{d}\Gamma d}{\sqrt{2}}\, , \nonumber \\
J_{f_0(a_0)}^2&=&\Sigma_{\Gamma=1,\pm \gamma_5} \bar{s}\Gamma
\frac{\lambda^a}{2} s\frac{\bar{u}\Gamma \frac{\lambda^a}{2} u\pm
\bar{d}\Gamma \frac{\lambda^a}{2} d}{\sqrt{2}},
\end{eqnarray}
where the $\lambda^a$ is the $SU(3)$ Gell-Mann matrix.   Perform
Fierz transformation both in the Dirac spinor and color space, for
example, we can obtain
 \begin{eqnarray}
J_{f_0}^2 &\propto& C_A
J^A_{f_0}+C_BJ^B_{f_0}+C_C\frac{\epsilon^{abc}\epsilon^{ade}}{\sqrt{2}}\left[(u_b^TC\gamma_\mu
s_c) (\bar{u}_d \gamma^\mu C\bar{s}_e^T)+(d_b^TC\gamma_\mu s_c)
(\bar{d}_d
\gamma^\mu C\bar{s}_e^T)\right] \nonumber \\
&&+C_D\frac{\epsilon^{abc}\epsilon^{ade}}{\sqrt{2}}\left[(u_b^TC\gamma_\mu
\gamma_5s_c) (\bar{u}_d \gamma^\mu
\gamma_5C\bar{s}_e^T)+(d_b^TC\gamma_\mu \gamma_5s_c) (\bar{d}_d
\gamma^\mu \gamma_5C\bar{s}_e^T)\right]\cdots \, , \nonumber \\
J_{a_0}^2 &\propto& C_A
J^A_{a_0}+C_BJ^B_{a_0}+C_C\frac{\epsilon^{abc}\epsilon^{ade}}{\sqrt{2}}\left[(u_b^TC\gamma_\mu
s_c) (\bar{u}_d \gamma^\mu C\bar{s}_e^T)-(d_b^TC\gamma_\mu s_c)
(\bar{d}_d
\gamma^\mu C\bar{s}_e^T)\right] \nonumber \\
&&+C_D\frac{\epsilon^{abc}\epsilon^{ade}}{\sqrt{2}}\left[(u_b^TC\gamma_\mu
\gamma_5s_c) (\bar{u}_d \gamma^\mu
\gamma_5C\bar{s}_e^T)-(d_b^TC\gamma_\mu \gamma_5s_c) (\bar{d}_d
\gamma^\mu \gamma_5C\bar{s}_e^T)\right]\cdots \, .
 \end{eqnarray}
 Here $C_A$, $C_B$, $C_C$ and $C_D$ are coefficients which are not shown explicitly for simplicity.
 In the color superconductivity theory,   the one gluon exchange induced  Nambu--Jona-Lasinio like Models
 will also lead to the $S^a-\bar{S}^a$ type and $P^a-\bar{P}^a$ type diquark
 pairs \cite{Cahill89},
 \begin{equation}
  G \bar{q}\gamma^\mu \frac{\lambda^a}{2} q \bar{q}\gamma_\mu \frac{\lambda^a}{2}
  q \propto C_A S^a \bar{S}^a + C_B P^a \bar{P}^a+\cdots \, .
 \end{equation}   So we can  take the point of view that  the lowest
lying scalar mesons are $S$-wave bound states of
diquark-antidiquark pairs of  $S^a-\bar{S}^a$ type and $P^a-\bar{P}^a$
type.

In this article, we investigate the masses of the scalar mesons
 $f_0(980)$ and $a_0(980)$ with two interpolating currents
 respectively and choose the following two-point correlation functions,
\begin{eqnarray}
\Pi_S^i(p)=i\int d^4x ~e^{ip.x}\langle 0
|T[J_S^i(x){J^i_S}^\dagger(0)]|0\rangle.
\end{eqnarray}
Here the  current  $J_S^i$ denotes $J^A_{f_0}$, $J^B_{f_0}$,
$J^A_{a_0}$ and $J^B_{a_0}$. According to the basic assumption of
current-hadron duality in the QCD sum rules  approach
\cite{Shifman79}, we insert  a complete series of intermediate
states satisfying the unitarity   principle with the same quantum
numbers as the current operator $J_S^i(x)$
 into the correlation functions in
Eq.(9)  to obtain the hadronic representation. Isolating the
ground state contributions from the pole terms of the mesons $f_0(980)$
and $a_0(980)$ , we get the  result,
\begin{eqnarray}
\Pi_S^i(p)=\frac{2f_S^{i2}m_s^{i8}}{m_S^{i2}-p^2}+\cdots \, ,
\end{eqnarray}
where the following definitions have been used,
\begin{equation}
 \langle 0 | J_S^i|S\rangle =\sqrt{2}f_S^im^{i4}_S \;.
 \end{equation}
We have not shown the contributions from the higher resonances and
continuum states explicitly for simplicity.

The  calculation of  operator product expansion in the  deep
Euclidean space-time region is
  straightforward and tedious, technical details are neglected for
  simplicity.  In this article, we consider the vacuum condensates up to dimension six.
  Once  the analytical  results are obtained,
  then we can take the current-hadron dualities below the thresholds
$s_0$ and perform the Borel transformation with respect to the
variable $P^2=-p^2$, finally we obtain  the following sum rules,
\begin{eqnarray}
2f^{A2}_{f_0(a_0)}m^{A8}_{f_0(a_0)}e^{-\frac{m_{f_0(a_0)}^{A2}}{M^2}}=AA , \\
2f^{B2}_{f_0(a_0)}m^{B8}_{f_0(a_0)}e^{-\frac{m_{f_0(a_0)}^{B2}}{M^2}}=BB ,
\end{eqnarray}
\begin{eqnarray}
AA&=&\int_{4m_s^2}^{s_0}ds e^{-\frac{s}{M^2}}\left\{
\frac{s^4}{2^9 5! \pi^6}+\frac{\langle \bar{s}s\rangle\langle
\bar{q}q\rangle s}{12\pi^2} +\frac{3\langle \bar{q}g_s \sigma G
q\rangle-\langle
\bar{s}g_s \sigma  G s\rangle}{2^6 3 \pi^4}m_s s \right. \nonumber\\
&&\left.-\frac{2\langle \bar{q} q\rangle-\langle \bar{s}
s\rangle}{2^6 3 \pi^4}m_s s^2 +\frac{s^2}{2^9 3 \pi^4} \langle
\frac{\alpha_s GG}{\pi}\rangle \right\}  , \nonumber \\
BB&=&\int_{4m_s^2}^{s_0}ds e^{-\frac{s}{M^2}}\left\{
-\frac{s^4}{2^9 5! \pi^6}+\frac{\langle \bar{s}s\rangle\langle
\bar{q}q\rangle s}{12\pi^2} +\frac{3\langle \bar{q}g_s \sigma  G
q\rangle+\langle
\bar{s}g_s \sigma  G s\rangle}{2^6 3 \pi^4}m_s s \right. \nonumber\\
&&\left.-\frac{2\langle \bar{q} q\rangle+\langle \bar{s}
s\rangle}{2^6 3 \pi^4}m_s s^2 -\frac{s^2}{2^9 3 \pi^4} \langle
\frac{\alpha_s GG}{\pi}\rangle \right\} .  \nonumber
\end{eqnarray}
Differentiate  the above sum rules with respect to the variable
$\frac{1}{M^2}$, then eliminate the quantities $f^A_{f_0(a_0)}$
and $f^B_{f_0(a_0)}$ , we obtain
\begin{eqnarray}
m^{A2}_{f_0(a_0)}&=&\int_{4m_s^2}^{s_0}ds
e^{-\frac{s}{M^2}}\left\{ \frac{s^5}{2^9 5! \pi^6}+\frac{\langle
\bar{s}s\rangle\langle \bar{q}q\rangle s^2}{12\pi^2}
+\frac{3\langle \bar{q}g_s \sigma  G q\rangle-\langle
\bar{s}g_s \sigma  G s\rangle}{2^6 3 \pi^4}m_s s^2 \right. \nonumber\\
&&\left.-\frac{2\langle \bar{q} q\rangle-\langle \bar{s}
s\rangle}{2^6 3 \pi^4}m_s s^3 +\frac{s^3}{2^9 3 \pi^4} \langle
\frac{\alpha_s GG}{\pi}\rangle \right\}/AA, \\
m^{B2}_{f_0(a_0)}&=&\int_{4m_s^2}^{s_0}ds
e^{-\frac{s}{M^2}}\left\{ -\frac{s^5}{2^9 5! \pi^6}+\frac{\langle
\bar{s}s\rangle\langle \bar{q}q\rangle s^2}{12\pi^2}
+\frac{3\langle \bar{q}g_s \sigma  G q\rangle+\langle
\bar{s}g_s \sigma  G s\rangle}{2^6 3 \pi^4}m_s s^2 \right. \nonumber\\
&&\left.-\frac{2\langle \bar{q} q\rangle+\langle \bar{s}
s\rangle}{2^6 3 \pi^4}m_s s^3 -\frac{s^3}{2^9 3 \pi^4} \langle
\frac{\alpha_s GG}{\pi}\rangle \right\} /BB.
\end{eqnarray}
 It is easy to perform the   $s$ integral in
 Eqs.(12-15),  we prefer this form for simplicity.

\section{Numerical Results}
The parameters are taken as $\langle \bar{s}s \rangle=0.8\langle
\bar{u}u \rangle$, $\langle \bar{s}g_s\sigma  G s
\rangle=0.8\langle \bar{s}s \rangle$, $\langle \bar{q}g_s\sigma
 G q \rangle=0.8\langle \bar{q}q \rangle$, $\langle \bar{u}u
\rangle=\langle \bar{d}d \rangle=\langle \bar{q}q \rangle=(-219
MeV)^3$, $\langle \frac{\alpha_sGG}{\pi} \rangle=(0.33 GeV)^4$,
 $m_u=m_d=0$ and $m_s=150MeV$.  The main contributions to the sum rules
 come from  the quark condensates terms, here we have taken the standard
values and neglected the uncertainties, small variations of those
condensates will not  lead to larger changes about   the numerical
 values.  The threshold parameter $s_0$ is chosen to
  vary between $(1.4-1.6) GeV^2$ to avoid possible pollutions from
  higher resonances and continuum states.
In the region $M^2=(1.2-3.2)GeV^2$, the sum rules for
$m^A_{f_0}=m^A_{a_0}$ and $m^B_{f_0}=m^B_{a_0}$  are almost
independent of  the Borel parameter $M^2$ which are plotted in the
Figure for $s_0=1.5GeV^2$ as an example.
  Due to the special quark constituents and Dirac structures of the
  interpolating currents, the scalar mesons $f_0(980)$ and $a_0(980)$ have degenerate masses.
   For the $S^a-\bar{S}^a$ type interpolating currents $J^A_{f_0}$ and $J^A_{a_0}$,  the values for
   masses are about $m^A_{f_0}=m^A_{a_0}=(0.96-1.02) GeV$,
  while for the $P^a-\bar{P}^a$ type interpolating currents $J^B_{f_0}$ and $J^B_{a_0}$,  the values for
   masses are about $m^B_{f_0}=m^B_{a_0}=(0.95-1.01) GeV$.
  Although the  values for masses $m^A_{f_0}=m^A_{a_0}$ lie a little above the masses $m^B_{f_0}=m^B_{a_0}$,
  we can not get to the conclusion that the scalar mesons $f_0(980)$ and $a_0(980)$
  prefer the $S^a-\bar{S}^a$ type interpolating currents $J^A_{f_0}$ and $J^A_{a_0}$
  to the $P^a-\bar{P}^a$ type interpolating currents $J^B_{f_0}$ and $J^B_{a_0}$.
Precise determination of what type interpolating currents we
should choose calls for original theoretical approaches,
 the contributions from  the direct instantons may do the work. In our recent work,
 we observe that the contributions from the  direct instantons are
considerable for the pentaquark state $\Theta^+(1540)$ \cite{Wang05}, furthermore,
 the contributions from the direct instantons can improve the QCD sum rule greatly
 in some channels,  for example,  the
nonperturbative contributions from the direct instantons to the
conventional operator product expansion can significantly improve
the stability of chirally odd  nucleon sum rules
\cite{Dorokhov90,Forkel}. Despite whatever the interpolating
currents may be, we observe that they both give the correct
degenerate masses for the scalar mesons $f_0(980)$ and $a_0(980)$,
there must be some four-quark constituents in those mesons.

\section{Conclusions}
In this article, we take the point of view that the $f_0(980)$ and
$a_0(980)$ mesons are  four-quark states
$(qq)_{\bar{3}}(\bar{q}\bar{q})_3$ in the ideal mixing limit, and
devote to determine the values of their masses $m_{f_0}$ and
$m_{a_{0}}$ in the framework of  the  QCD sum rules approach.  Due
to the special quark constituents and Dirac structures of the
  interpolating currents, the scalar mesons $f_0(980)$ and $a_0(980)$ have degenerate masses.
    For the $S^a-\bar{S}^a$ type interpolating currents $J^A_{f_0}$ and $J^A_{a_0}$,  the values for
   masses are about $m^A_{f_0}=m^A_{a_0}=(0.96-1.02) GeV$,
  while for the $P^a-\bar{P}^a$ type interpolating currents $J^B_{f_0}$ and $J^B_{a_0}$,  the values for
   masses are about $m^B_{f_0}=m^B_{a_0}=(0.95-1.01) GeV$.
  Although the  values for masses $m^A_{f_0}=m^A_{a_0}$ lie a little above the masses $m^B_{f_0}=m^B_{a_0}$,
  we can not get to the conclusion that the scalar mesons $f_0(980)$ and $a_0(980)$
  prefer the $S^a-\bar{S}^a$ type interpolating currents $J^A_{f_0}$ and $J^A_{a_0}$
  to the $P^a-\bar{P}^a$ type interpolating currents $J^B_{f_0}$ and $J^B_{a_0}$.
 Despite whatever the interpolating
currents may be, we observe that they both give the correct
degenerate masses for the scalar mesons $f_0(980)$ and $a_0(980)$,
there must be some four-quark constituents in those mesons, our
results support the four-quark model  and the  hybrid model.  In
the  hybrid model,  those mesons are
 four-quark states $(qq)_{\bar{3}}(\bar{q}\bar{q})_3$ in S-wave near the
center, with some constituent $q \bar{ q}$ in P-wave, but further out they
rearrange into  $(q \bar{ q})_1(q \bar{ q})_1$ states and finally
as meson-meson states \cite{Close2002}. Precise determination of what type
interpolating currents we should choose calls for original
theoretical approaches,
 the contributions from  the direct instantons may do the work.

\section*{Acknowledgment}
This  work is supported by National Natural Science Foundation,
Grant Number 10405009,  and Key Program Foundation of NCEPU. The
authors are indebted to Dr. J.He (IHEP) , Dr. X.B.Huang (PKU) and Dr. L.Li (GSCAS)
for numerous help, without them, the work would not be finished. The author would
also thanks Prof. M.Nielsen for helpful discussion.

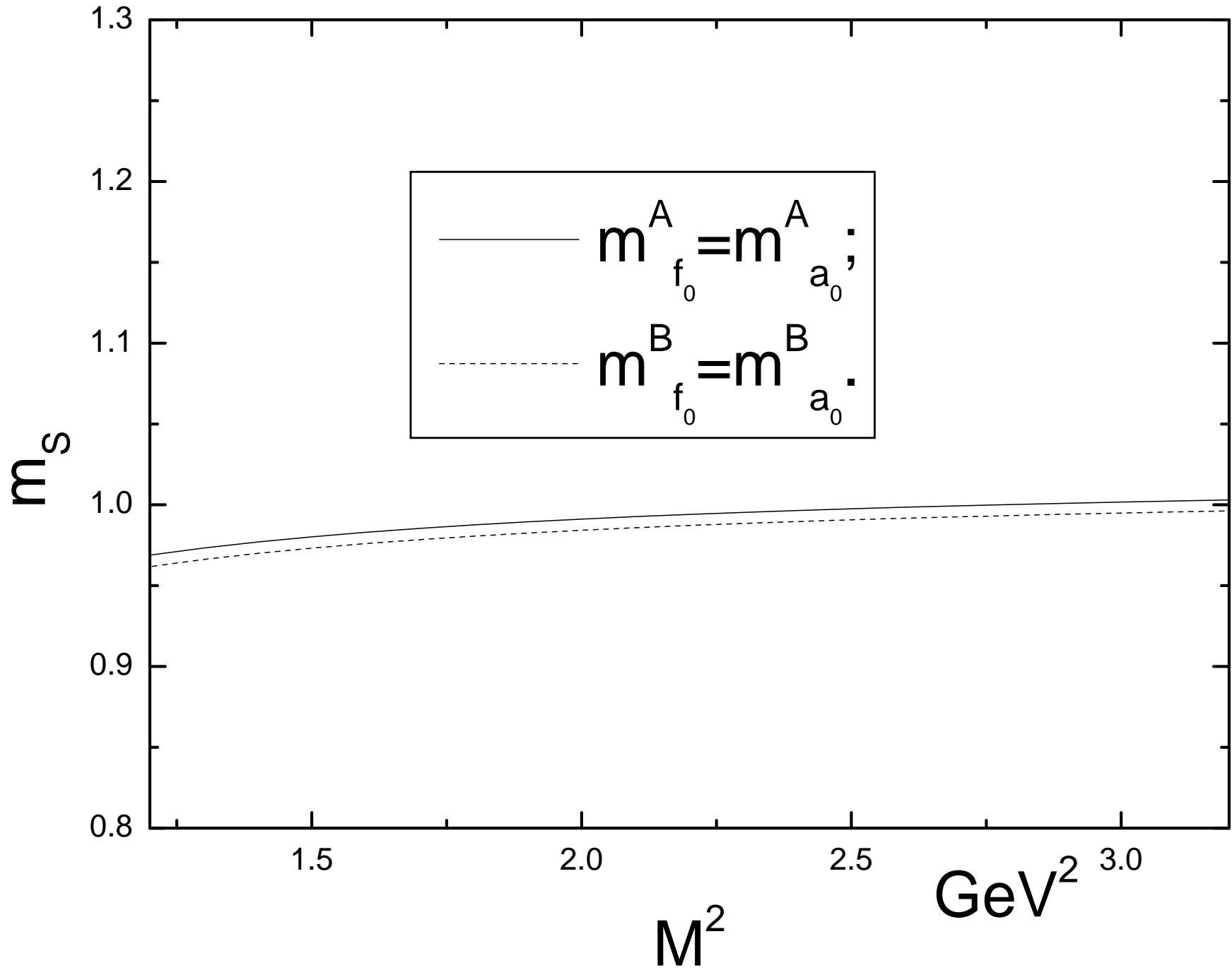
\end{document}